\documentstyle[11pt,paspconf,epsf]{article}

\begin{document}

\title{Evolution of Globular Clusters: Effects of Tidal Shocks}

\author{Oleg Y. Gnedin \& Jeremiah P. Ostriker}
\affil{Princeton University Observatory\\
       ognedin@astro.princeton.edu, jpo@astro.princeton.edu}

\begin{abstract}
The semi-analytic theory of tidal shocks proves to be a powerful tool
to study tidal interactions of star clusters and satellite galaxies
with their massive hosts.  New models of the globular cluster evolution
employ a combination of analytic estimates, solutions of the Fokker-Planck
equation and direct N-body simulations.  The models predict large destruction
rates for the Galactic globular clusters.  Those on the highly
eccentric orbits around the Galactic center are much more likely to
be disrupted than the ones on nearly circular orbits.
The destruction rates are largely increased near the bulge.
Disruption of the low-mass clusters changes the Luminosity
Function of the Globular Cluster System, shifting the peak
of the Luminosity Function to the brighter end.
\end{abstract}


\section{Introduction}

Dynamical evolution of globular clusters is strongly affected by
gravitational tidal shocks.  When the clusters cross the disk of
the Galaxy they experience disk shocking;  when the clusters pass
near the Galactic center, they experience bulge shocking.
The effects of the tidal shocks depend on the density of the background
stars and are especially pronounced in the inner regions of the
Galaxy.  Tidal shocks increase the energy of random motion of stars,
reduce the binding energy of the cluster, accelerate core collapse,
and lead to the faster overall evolution and destruction of globular
clusters (for example,  Spitzer 1987;  Weinberg 1994;
Murali \& Weinberg 1997a,b,c;  Gnedin \& Ostriker 1997a).

\section{Destruction of Globular Clusters}

We calculate the rate of destruction of globular clusters as a result
of various physical processes:  two-body relaxation, evaporation of
stars through the tidal boundary, disk shocking and bulge shocking
(Gnedin \& Ostriker 1997a).  We use a Fokker-Planck code
which includes tidal shocks semi-analytically.  We introduce the
adiabatic corrections that account for the conservation of adiabatic
invariants of the fast moving stars
(Gnedin \& Ostriker 1997b).
These corrections reduce the energy input due to the shocks in the inner
parts of the cluster.  The results (Figure 1) show that tidal shocks
dominate cluster evolution near the Galactic center.  Overall, as many as
50\% to 90\% of the present sample of globular clusters may be destroyed
within the next Hubble time.

\begin{figure}[t]
\vspace{8.5cm}
\includegraphics{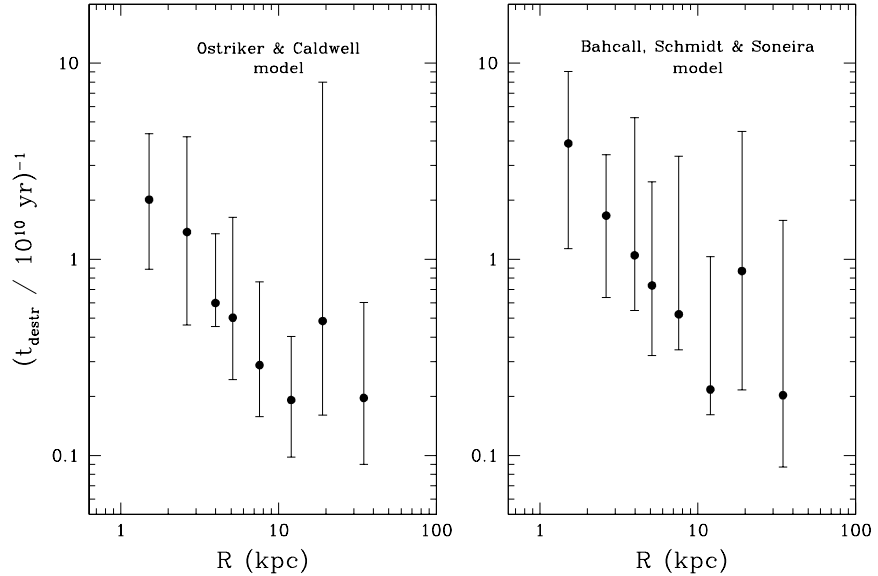}
\caption{
The destruction rates for the Galactic globular clusters, defined as the
inverse time to destruction in units of $10^{10}$ yr, versus their
present position in the Galaxy.  The two panels
show the results of the Fokker-Planck calculations for the two
Galactic models and the isotropic velocity distribution of globulars.}
\label{fig1}
\end{figure}

\begin{figure}[t]
\vspace{10cm}
\includegraphics{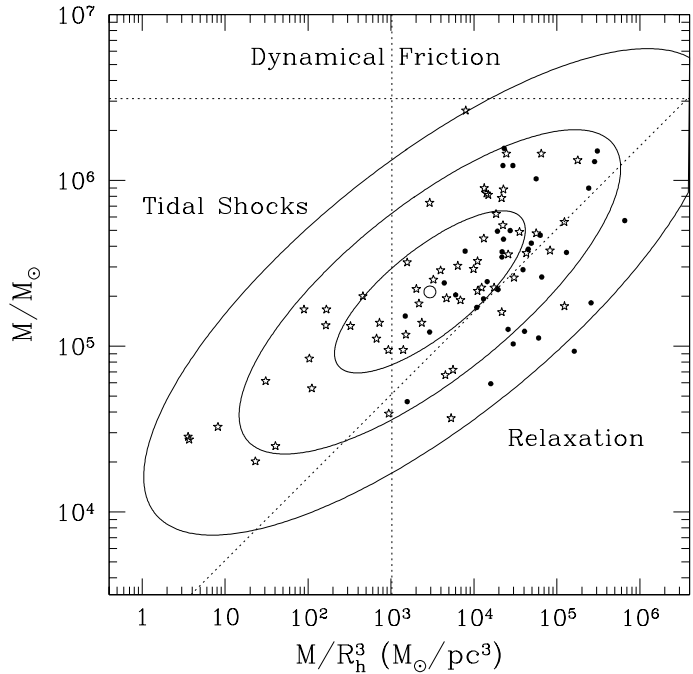}
\caption{
Distribution of the inner (green dots) and outer (red stars)
Galactic globular clusters on the $\log{M}-\log{\rho}$ plane.
The solid lines show the intrinsic distribution,
where the ellipses are $1\sigma$, $2\sigma$,
and $3\sigma$ levels, and the small circle is at the center
of the distribution. The dotted lines mark the region allowed for the
inner clusters by dynamical processes: above the horizontal line,
dynamical friction is important, to the left of the vertical line tidal
shocks are important, and below the diagonal line, relaxation will lead
to core collapse and subsequent disintegration of the clusters.}
\label{fig2}
\end{figure}

\begin{figure}[t]
\vspace{10cm}
\includegraphics{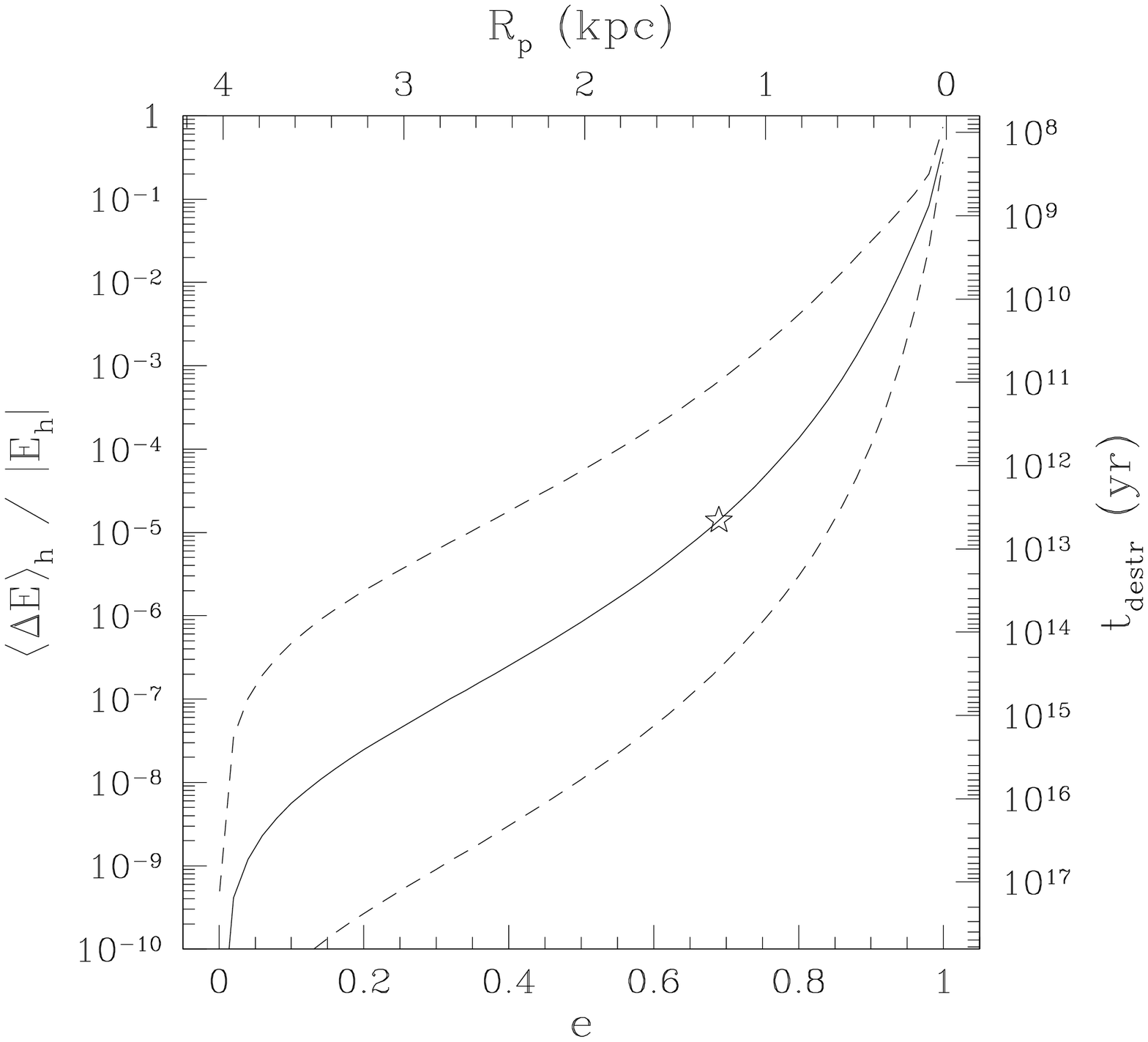}
\caption{
The energy change of stars at the half-mass radius of NGC 6712
relative to the initial half-mass energy as a function of eccentricity
of the orbit.  The orbital energy is fixed at the observed value for
the cluster (red line) or at the one-half and twice that value
(dashed green and blue lines, respectively).}
\label{fig3}
\end{figure}

\section{Evolution of the Luminosity Function}

Tidal shocks destroy most easily the low-mass, low-density clusters.
Figure 2 shows the distribution of the inner and outer
Galactic globular
clusters.  As expected, there are no low-density clusters in the inner
part of the Galaxy where the tidal shocks operate most efficiently.
Removal of those low-mass clusters makes the mean, or the peak, of the
Luminosity Function (LF) to shift towards bighter magnitudes.

An apparent correlation of the masses and densities of outer clusters
allows us to construct an intrinsic distribution of globulars, unaffected
by the shocks.  By applying the dynamical calculations, we can estimate
the amount of brightening of the peak of the LF.  Assuming that in all
galaxies the initial distribution is the same, we can reconstruct the
shock history in an external galaxy and infer the peak of the original
distribution of globular clusters.  Comparing that peak with the center
of the intrinsic distribution in the Galaxy, we obtain a distance
estimate to the galaxy.  This method is fully independent and makes
unnecessary the common assumption that the peak of the LF is a
standard candle.  Applied to
the best known samples of M31 and M87
(Ostriker \& Gnedin 1997),
our method gives a distance estimate in very close agreement
with that obtained with Cepheids and other methods.

\section{Example: NGC 6712}

Tidal heating can be simply parametrized to study semi-analytically the
evolution of the individual clusters.  We derive analytic equations for
the first and second order energy changes, $\langle \Delta E \rangle$
and $\langle \Delta E^2 \rangle$, of stars in the cluster
(Gnedin, Hernquist \& Ostriker 1997).
These equations are supplemented
by the adiabatic corrections depending on the effective duration of the
shock.  Heating on the nearly circular orbits is strongly suppressed
because the ``shock'' becomes very slow (Figure 3).
The analytical
estimates are tested against the Self-Consistent N-body simulations
of the shocking event along a true trajectory of the cluster in the
Galaxy.  We find a remarkable agreement with the
simulations.

\acknowledgments
This project was supported in part by NSF grant AST 94-24416.

\end{document}